\documentclass[12pt]{article}
\usepackage{amssymb}
\usepackage{psfig}
\begin{document}
\addtolength{\baselineskip}{.5mm}
\newlength{\extraspace}
\setlength{\extraspace}{1.5mm}
\newlength{\extraspaces}
\setlength{\extraspaces}{2mm}

\newcommand{\newsection}[1]{
\vspace{15mm} \pagebreak[3] \addtocounter{section}{1}
\setcounter{subsection}{0} \setcounter{footnote}{0}
\noindent {\Large\bf \thesection. #1} \nopagebreak
\medskip
\nopagebreak}

\newcommand{\newsubsection}[1]{
\vspace{1cm} \pagebreak[3] \addtocounter{subsection}{1}
\addcontentsline{toc}{subsection}{\protect
\numberline{\arabic{section}.\arabic{subsection}}{#1}}
\noindent{\large\bf 
\thesubsection. #1} \nopagebreak \vspace{2mm} \nopagebreak}
\newcommand{\ba}{\begin{eqnarray}
\addtolength{\abovedisplayskip}{\extraspaces}
\addtolength{\belowdisplayskip}{\extraspaces}
\addtolength{\abovedisplayshortskip}{\extraspace}
\addtolength{\belowdisplayshortskip}{\extraspace}}
\newcommand{\zbar}{\overline{z}}
\newcommand{\tr}{{\rm tr\,}}
\newcommand{\ea}{\end{eqnarray}}
\newcommand{\OL}[1]{ \hspace{2pt}\overline{\hspace{-2pt}#1
   \hspace{-1pt}}\hspace{1pt} }
\newcommand{\is}{& \! = \! &}
\newcommand{\eps}{\epsilon}
\newcommand{\calR}{{\cal R}}
\newcommand{\calB}{{\cal M}}
\newcommand{\calK}{{\cal K}}
\newcommand{\calG}{{\cal G}}
\newcommand{\tilF}{{\tilde F}}
\newcommand{\barG}{{\OL G}}
\newcommand{\alphap}{}
\newcommand{\be}{\begin{equation}
\addtolength{\abovedisplayskip}{\extraspaces}
\addtolength{\belowdisplayskip}{\extraspaces}
\addtolength{\abovedisplayshortskip}{\extraspace}
\addtolength{\belowdisplayshortskip}{\extraspace}}
\newcommand{\ee}{\end{equation}}
\newcommand{\STr}{{\rm STr}}
\newcommand{\figuur}[3]{
\begin{figure}[t]\begin{center}
\leavevmode\hbox{\epsfxsize=#2 \epsffile{#1.eps}}\\[3mm]
\parbox{15.5cm}{\small
\it #3}
\end{center}
\end{figure}}
\newcommand{\im}{{\rm Im}}
\newcommand{\calm}{{\cal M}}
\newcommand{\call}{{\cal L}}
\newcommand{\sect}[1]{\section{#1}}
\newcommand\hi{{\rm i}}
\newcommand\MMs{{\mbox{\small $M$}}_{\! s}}
\newcommand\MMp{{\mbox{\small $M$}}_{\! p}}

\newcommand{\ANOT}{A_0}
\newcommand{\anot}{a_0}

\begin{titlepage}
\begin{center}

{\hbox to\hsize{ \hfill PUPT-2114}}
{\hbox to\hsize{ \hfill SLAC-PUB-10326}}
{\hbox to\hsize{ \hfill SU-ITP-04/03}}
{\hbox to\hsize{ \hfill hep-th/0403123}}

\vspace{3.5cm}

{\Large \bf The Giant Inflaton}
\\[1.5cm]

{Oliver DeWolfe${}^1$, Shamit Kachru${}^2$ and Herman Verlinde${}^1$}\\[8mm]

{${}^1$ \it Department of Physics, Princeton University,
Princeton,
NJ 08544}\\[1mm]
{${}^2$ \it Department of Physics and SLAC, Stanford University,
Stanford, CA 94305/94309}

\vspace*{2.5cm}

{\bf Abstract}\\

\end{center}
\noindent We investigate a new mechanism for realizing slow roll
inflation in string theory, based on the dynamics of $p$ anti-D3
branes in a class of mildly warped flux compactifications. 
Attracted to the bottom of a warped conifold throat, the anti-branes then cluster due
to a novel mechanism wherein the background flux polarizes in an attempt to screen them. Once they are sufficiently close, the $M$ units of flux cause the anti-branes to expand into a fuzzy NS5-brane,
which for rather generic choices of $p/M$ will unwrap around the geometry, decaying into
D3-branes via a classical process.   We find that the effective potential governing this
evolution possesses several epochs that can potentially support slow-roll inflation, provided
the process can be arranged to take place at a high enough energy scale, of about one or two
orders of magnitude below the Planck energy; this scale, however, lies just outside the bounds of our approximations.

\end{titlepage}

\newcommand{\Tr}{{\rm Tr}}

\newpage
\newsection{Introduction}

Inflation is presently the most attractive scenario for early
cosmology \cite{infclass}. The assumption that the universe has
gone through an early de Sitter phase, driven by a slowly rolling
inflaton field, naturally predicts a flat universe and can produce
a nearly scale-invariant spectrum of density perturbations, in
agreement with current observations. In a successful inflation
model, however, the inflaton potential must be quite delicately
tuned to satisfy various constraints: it must be sufficiently flat
to produce at least 60 e-foldings of expansion, it must allow for
a graceful exit from inflation, and there must be a natural
mechanism for reheating and producing density perturbations of the
correct magnitude. It is therefore important to know whether
realistic models of inflation can naturally arise from a
microscopic starting point such as string theory.

To obtain a string realization of inflation, one preferably would
like to start from a string compactification with fixed shape,
size, and string coupling, since experience shows that when
unfixed these moduli typically have too steep a potential to
permit inflation. Finding such stable compactifications is an
important but difficult problem. Promising scenarios for
stabilizing all geometric moduli have recently been discussed
within the context of warped type IIB flux compactifications in
\cite{KKLT,BKQ,AlexEva,UCSB,Quev,Brustein}.  These flux compactifications
have several other features that make them attractive starting
points for constructing string inflation models. The geometrical
warping in these scenarios can provide a dynamical mechanism to
control the size of potentially destabilizing
supersymmetry-breaking effects, by introducing a hierarchy of
scales.  Most relevant for us, they naturally incorporate mobile
branes.

When anti-branes are introduced, their tension can provide the requisite positive vacuum energy necessary for inflation.  Furthermore, as we shall make explicit in this note, fields corresponding to their positions on the compact space can naturally possess a sufficiently flat potential to be candidate inflatons.  One then requires a graceful exit mechanism, a classical process by which the vacuum energy stored in the anti-branes can decay.

In typical brane inflation scenarios considered thus far
\cite{bantib}, one imagines an inflationary system with both
$\overline {\rm D3}$- and D3-branes. The brane/anti-brane distance
is the candidate inflaton, and the exit from inflation takes place
via the violent brane/anti-brane annihilation process. The
embedding of such inflationary models in warped flux
compactifications was studied in detail in \cite{KKLMNOPQRST}. The
conclusion was that, either due to the mutual attraction between
the branes or due to coupling with the K\"ahler moduli, the
potential in such a model is generally too steep to support
inflation.

We shall consider a different, more ``stringy'' exit
scenario, which has the advantage that it only requires anti-branes.
As shown in \cite{KPV}, it is possible for $\overline {\rm D3}$'s in a warped
flux geometry (such as the example of the Klebanov-Strassler throat \cite{KS}) to
annihilate against the background flux, via the intermediate
formation of a ``giant graviton" 5-brane. Moreover, it was found that
for a sufficient number of $\overline {\rm D3}$'s, this decay
proceeds as a classical (as opposed to quantum tunneling)
process, and thus could represent a viable exit mechanism for
inflation.

Taking this decay as a proposal for an exit from inflation, we consider the dynamics of a number of $\overline {\rm D3}$'s as they evolve towards it.  As we shall discuss, there are several distinct phases in the evolution that may be able to support a slow roll phase.  In this paper, we systematically examine these phases in the brane life cycle as possible inflationary epochs.

We begin by simply placing a number $p$ of $\overline {\rm D3}$'s  inside a stabilized flux compactification (the details of the  stabilization do not matter much for us here).  As in \cite{GKP}, we assume that the geometry includes a (mildly) warped conifold
region \cite{KS}. The $\overline {\rm D3}$'s will automatically be
drawn down the ``throat" towards the $S^3$ at the tip of the conifold.  Although $\overline{\rm D3}$'s feel no force from one another in flat space, this is not the case in the flux geometry.
We demonstrate an interesting mechanism wherein the fluxes are polarized in an attempt to screen the anti-branes, and the anti-branes then feel a force from the inhomogeneous background.  The first stage in the evolution is hence that the anti-branes begin to cluster together.  When they come close enough to one
other, the Myers effect \cite{Myers} takes over as in \cite{KPV},
and their worldvolume scalars condense to form a coherent
non-Abelian configuration, an NS-5 brane that we christen the giant inflaton.
The dynamics of giant graviton formation is a stringy effect not occurring in most brane world models, relying on the appearance of non-Abelian gauge theory when the branes
coincide and the detailed interactions of worldvolume scalars with
the background flux. When enough anti-branes have coalesced into a single giant, the 5-brane becomes able to
unwrap itself by traversing the $S^3$, finally decaying and depositing all its
potential energy into the matter that lives on a
newly created set of (supersymmetric) D3-branes.

For a suitable choice of parameters, we find that all three
stages, the accumulation of the anti-branes, the giant inflaton
formation, and the unwrapping process, can lead to a substantial
amount of inflation, provided the string scale at the bottom of
the conifold can be chosen high enough.   This  condition, however, implies a rather strict lower bound on the amount of warping, and our approximations become less reliable in this regime.  
Hence although the scenario has some promising features, it eludes a precise, controllable realization.

Because none of the potential inflationary stages involve motion in the
radial direction of the throat, these scenarios can evade the
problems arising from the conformal coupling in AdS-like regions
of warped geometries \cite{KKLMNOPQRST}.

This paper is organized as follows. In \S2 we start with an overview of the
various stages of our inflationary model. The various stages are
then considered in quantitative detail in \S3, \S4 and \S5. Each
section ends with an estimate of the conditions necessary for
inflation, which will depend on the ratio the 4-d Planck scale and
the string scale at the bottom of the warped geometry. In \S6, we
estimate this ratio, finding that the conditions for inflation to
occur -- namely very mild warping -- may be just outside the
regime of validity of our approximations. We close with a summary
of some general lessons from this work in \S7. Some calculations
which are referred to in the body of the paper but whose details
are not essential are relegated to appendices.

While this work was in progress, an idea which is similar in
spirit but not in detail, appeared in the paper of Pilo, Riotto
and Zaffaroni \cite{PRZ}. Other promising recent work concerning
stringy inflation models can be found in \cite{recentgo, Evatong, BR,  Garousi, Buchel, Cline}, where different ideas for overcoming the difficulties described in \cite{KKLMNOPQRST} are
discussed.

\newsection{The Life Cycle of the Anti-D3 Brane}

We begin with an overview of the dynamics experienced by a set of $\overline{\rm D3}$-branes on the road towards giant graviton decay, and highlight the epochs in which slow roll inflation seems possible.  This section also serves as an introduction and summary of the subsequent three sections.

\newsubsection{Setting: Warped Flux Compactification}

Our inflationary scenario is realized within a warped
compactification of type IIB string theory to four dimensions. We
briefly review the warped backgrounds, following \cite{GKP}; our
conventions are those of \cite{PS}. We work in string units $\alpha'=1$.
The full geometry has the form
\begin{eqnarray}
ds^2 = e^{2A} g_{\mu\nu} dx^\mu dx^\nu +  e^{-2A} \tilde{g}_{mn}
dy^m dy^n \,,
\end{eqnarray}
where $e^A(y)$ is the warp factor and $g_{\mu\nu}$ is the 4D metric.
The unwarped compact metric $\tilde{g}_{mn}$ is that of
a Calabi-Yau threefold.\footnote{In the F-theory generalization,
non-constant axion/dilaton fields require a non-Calabi-Yau
background 6-geometry, though the data of the geometry along with
the varying axio-dilaton is summarized by a Calabi-Yau fourfold.}
The geometry is additionally threaded by three- and five-form field strengths. The five-form $\tilde{F}_5$ is self-dual in 10 dimensions, and is given by
\begin{eqnarray}
\tilde{F}_5 = {\cal F}_5 + * {\cal F}_5 \,, \quad \quad {\cal F}_5
=
 d \alpha \wedge dV_4 / g_s \,,
\end{eqnarray}
for some function $\alpha(y)$, where $dV_4 =  \sqrt{-g_4} \, d^4
x$. The RR and NSNS three-form field strengths $F_3$ and $H_3$ are conveniently assembled into the complex combination
\begin{eqnarray}
\label{gthree}
G_3 \equiv F_3 - \tau H_3 \,,
\end{eqnarray}
where $\tau = C_0 +i e^{-\phi}$ is the axion-dilaton.

Three-form fluxes with support on given three cycles of the
Calabi-Yau manifold generate a warp factor and fix the complex
structure moduli \cite{GKP}. Depending on the choice of fluxes,
this may result in one or more conical regions with an AdS-like
geometry. We will primarily be concerned with dynamics in a single
warped throat with $M$ units of $F_{3}$ flux through the $A$-cycle
and $K$ units of $H_{3}$ through the dual $B$-cycle:
\begin{eqnarray}
\label{3FormFlux} {1 \over 2 \pi} \int_A F_3 = 2 \pi M \,, \quad
\quad {1 \over 2 \pi} \int_B H_3 = - 2 \pi K \,,
\end{eqnarray}
where
$M$ and $K$ are integers. To simplify our discussion, we will
assume that $M$ and $K$ are the only crossed three-form fluxes
that are turned on.

Besides flux, the geometry will typically involve the insertion of
$N_3$ D3-branes and/or $\overline{N}_3$ anti-D3 branes, localized
at points in the compact space. The net 5-form charge is required
to vanish by the integrated Bianchi identity, leading to the condition
\cite{tadpole}
\begin{equation}
\label{tadpole} {\chi(X)\over 24} ~=~  Q_{3}
+ M K \,. \label{tadpoleF}
\end{equation}
Here $Q_{3} = N_3 - \overline{N}_3$ is the net charge from mobile
branes.  The Euler characteristic $\chi(X)$ of the F-theory CY
fourfold gives the net charge from 7-branes wrapped on 4-cycles;
for us $\chi(X)$ can be thought of as a property of the background
providing a sink to absorb the charge on the RHS of
(\ref{tadpole}). The typical value of $\chi(X)$ can be quite
large; it is easy to find examples in which $\chi(X)/24$ is of
order $10^4$ or larger. Hence if we choose $K$ relatively small,
we can consider values for $M$ of up to $10^3$ or even larger.

When $\overline{\rm D3}\,$'s are absent, there exist certain
special warped backgrounds over flat four-dimensional space, where
the fluxes are imaginary self-dual  (ISD)  \cite{fcomp,wcomp} and
the warp factor is related to the 5-form flux:
\begin{eqnarray}
\label{ISD} *_6 G_3 = i G_3 \,, \quad \quad e^{4A} = \alpha \,.
\end{eqnarray}
The imaginary self-duality condition requires $G_3$ to have
contributions only from $(2,1)$ and $(0,3)$ indices relative to
the complex structure; the former preserves supersymmetry while
the latter breaks it.  These solutions have been termed
``pseudo-BPS" because despite the fact that supersymmetry may be
broken, mobile D3-branes feel no force from the background or each
other,\footnote{This lack of force may be modified by the
volume-stabilization mechanism.} and their backreaction does not
spoil the structure.

The fluxes and branes act as sources for the warp factor:
\begin{eqnarray}
\label{EinsteinEq} \nabla^2 A= {g_s \, G_{mnp} \overline{G}^{mnp}
\over 48 \, {\rm Im} \, \tau} + {e^{-8A} \over 4} \left(
\partial_m \alpha \partial^m \alpha - \partial_m e^{4A} \partial^m
e^{4A} \right) - e^{-2A} {\cal R}_4 + {\rm branes} \,,
\end{eqnarray}
where the warped metric $g_{mn}$ is used, and we have included the
term from the 4-dimensional Ricci scalar ${\cal R}_4$.  When the fluxes (\ref{3FormFlux})
$M$ and $K$ are defined on the $A$- and $B$-cycles of a conifold singularity within the total space, they generate an $AdS$-like warped throat coming to a smooth end, of the type studied by Klebanov and Strassler (KS) \cite{KS}; this throat, and its tip in particular, will be the arena for our inflation scenario.

As emphasized in \cite{KKLMNOPQRST}, the overall volume must also be stabilized to prevent the anti-branes from triggering a runaway decompactification.  We assume that the volume is somehow stabilized, though our discussion does not require any particular mechanism.\footnote{We note, however, that were the volume to be stabilized by the mechanism of \cite{KKLT}, our scenario does not encounter the problems found in \cite{KKLMNOPQRST} coming from the form of the K\"ahler potential \cite{DG}, as the motion we are interested in is exclusively along the equiK\"ahlerpotential at the bottom of the throat.}

\newsubsection{The Four Stages}

Now consider the case where only anti D3-branes are present in the geometry,
\begin{eqnarray}
\label{ourcase} N_3=0, ~{\overline
N_3}=p,~{\chi\over 24} ~=~KM - p \,,
\end{eqnarray}
This theory is
non-supersymmetric because the supersymmetry preserved by the
${\overline {\rm D3}}$'s is incompatible with the global
supersymmetry preserved by the ISD 3-form flux.  We assume that $p
\ll KM$ so that we may neglect the backreaction of the antibranes
on the background, except in a small neighborhood of the branes
themselves. Initially, the $p$ $\overline{\rm D3}$'s are placed at
random positions over the 6-d compactification manifold. In the
following, we will describe their subsequent life story. Our
discussion is based on their worldvolume action, which for a
non-essential technical reason we prefer to write in the S-dual
frame. It is given by \begin{equation} \label{nbi3} S_{\overline{\rm D3}} = - {\mu_3\over g_s}
\int d^4x \Tr\sqrt{\det( G_\parallel) \det( Q )} -  \mu_3 \int {\rm
Tr}\, (2 \pi i \, {\bf i}_\Phi {\bf i}_\Phi B_{\it 6} + C_4) \,, \end{equation}
where  $G_\parallel$ is the pullback of the induced metric along the brane, $\mu_3$ is the brane tension,  ${\bf i}_\Phi$ is the interior derivative, ${\bf i}_\Phi {\bf i}_\Phi B_6 = \Phi^n \Phi^m B_{mnpqrs} dy^p \wedge \ldots \wedge dy^s / 4!$, and
\begin{eqnarray}
\label{qqq}
Q^i{}_j = \delta^i{}_j + {2\pi i \over g_s}\, [ \Phi^i, \Phi^k]\,
(G_{kj}+g_s C_{kj})\,.
\end{eqnarray}
The scalar fields $\Phi \equiv 2 \pi X$ parameterize the location $X$ of the
${\overline{\rm D3}}$ branes.

\bigskip
\bigskip

\noindent {\sc Stage 0: Motion towards Apex}

In the very first stage of their life,
the anti-D3 branes are quickly drawn towards the region with the
smallest value of the warp-factor. This is seen as follows.

Let us introduce a coordinate system such that the warp factor depends
on some ``radial'' coordinate $r$ going down the throat.
The basic non-commutator terms of the worldvolume action of the
anti-branes in the ISD background are
\be - {\mu_3
\over g_s} \int d^4x \sqrt{g_4} \, {\Tr}\, e^{4A} \left(2 + {1
\over 2} e^{-2A} \partial^\mu \Phi^i \partial_\mu \Phi^j g_{ij} \right) \,,
\ee
The potential $V \sim 2 e^{4A}(r)$ comes from a combination of
Born-Infeld and Chern-Simons terms that cancel in the D3-brane case.
It generates a radial
force, $F_r(r)$ \begin{eqnarray} \label{force} {F}_r(r)
= -{2\mu_3\over g_s} \partial_r e^{4A}(r) \,, \end{eqnarray}
pulling the $\overline{D3}$-branes to the region of with the smallest value of the  warp factor: the tip of the conifold geometry.

\bigskip
\bigskip

\noindent{\sc New Setting: Geometry at the Apex}

In the following we will therefore assume that all of the interesting
dynamics takes place very close to the tip of the conifold; here we give a  brief description of this region. The
metric near the apex takes the form \cite{remarks}
\begin{eqnarray} \label{apex} ds^2  \simeq
 \anot^2\;
dx_\mu dx_\mu
+  \, R^2 d\Omega_3^2 + dr^2   + r^2 d\tilde{\Omega}_2^2\, .
\end{eqnarray}
The geometry of the tip $r=0$ is well approximated by a
three-sphere, with radius
\begin{eqnarray}
R^2 \simeq g_s M \,,
\end{eqnarray}
with $M$ the three-form RR-flux through the $S^3$ (\ref{3FormFlux}).  The conifold geometry has an $SO(4)$ symmetry acting naturally on the $S^3$ at the base of the throat. The embedding of the throat region
into the compact CY will break this symmetry, however.  To the extent that the $SO(4)$ is preserved, the RR three-form locally takes the form
\begin{eqnarray}
\label{flux1}
F_{mnp} = f \epsilon_{mnp} \,, \quad \quad f \, \simeq {2 \over \sqrt{g^3_s
M}}\,,
\end{eqnarray}
where $\epsilon_{mnp}$ is the warped volume element on the $S^3$.  In addition there is an NS three-form flux $H_3$, which due to the imaginary self-duality condition (\ref{ISD}) obeys $*_6 H_3 = -g_s F_3$.

The prefactor $\anot \equiv e^A|_{apex}$ in (\ref{apex}) is the value of the warp factor at the apex: it represents the redshift factor between the bulk of the CY geometry and  the tip of the conifold. Depending on the choice of fluxes
$K$ and $M$, it can be tuned to take an exponentially small value
\cite{GKP}. However, since the physics that could lead to
inflation takes place at the tip, we will in fact not be interested
in generating a large hierarchy between this scale and the Planck
scale;  instead, we will be drawn to a compactification scenario with
only mild warping. We will return to the physics of the
warp factor in \S6, where we will discuss the inflationary
parameters of our model. For now, we will treat $\anot$ as an
independently tunable quantity.

\bigskip
\bigskip

\noindent {\sc Stage I: Mutual Attraction}

The next stage starts with the $p$ anti-D3 branes scattered
randomly over the $S^3$ at the tip of the conifold. Since
anti-branes in flat space do not feel a force from one another,
and since the $S^3$ has an approximate $SO(4)$-symmetry, it would
seem a reasonable hope that the individual brane positions
$\Phi$ are like pseudo-Goldstone bosons, associated with spontaneous
breaking of the $SO(4)$-symmetry that acts on each brane-position.
In this case, the brane positions would be good candidates for inflaton
fields. In the compact background with three form flux, however,  the
anti-branes break the supersymmetry of the background, and one may
naturally wonder whether any additional force arises.  There will be two mechanisms
that concern us.

Although the KS-type throat respects the $SO(4)$ symmetry, the full CY geometry need not, and
consequently it will in general produce an effective potential on the $S^3$ that is common for every 3-brane.
For example, we expect to have to turn on at least one more flux in
order to stabilize the dilaton, as is described in \cite{GKP}, and
this flux will generically  be a source of $SO(4)$ symmetry
breaking.  The magnitude of the symmetry breaking from such
``distant" fluxes will however be suppressed by the warp factor, and will be small compared with the effects we discuss next. We
present the calculation of these forces in appendix A, from both
a direct supergravity perspective and a holographic field theory
perspective.

The second, more important effect that we need to include comes
from an effective mutual interaction that is induced between the branes.
This interaction is not suppressed by the warp-factor $\anot$, since it is
generated by local physics near the $S^3$. Still, it would seem a
reasonable hope that any such force vanishes at least at linearized order.
Somewhat surprisingly, as we will show in \S3, it turns
out that an interbrane force is already generated at the
linearized level.

The underlying mechanism is quite interesting: the branes polarize the surrounding flux background.  The background three-form fluxes have effective D3-brane charge, as is evident from (\ref{tadpole}), and they adjust themselves in an attempt to screen the anti-branes. As a result, the gravitational interaction dominates, producing an attractive
force between the anti-branes.  Equivalently, a probe anti-D3 ignores the other anti-branes but is drawn to the cloud of flux that is induced around them.  The typical magnitude of the force is comparable to that between a brane and an anti-brane. As a result the anti-branes will accumulate,  forming a single cluster.

We are led to ask whether this force can be weak enough that the branes can roll slowly as they come together.  We end \S3 by examining the condition for inflation during the
accumulation process.

\bigskip

\bigskip

\noindent{\bf\sc Stage II: Formation of the Non-Abelian Giant Inflaton}

If the branes are close to one another, by making their matrix
coordinates non-commutative, they can collectively represent a
5-dimensional brane which can be identified with the NS 5-brane \cite{KPV}.
The topology of this ``fuzzy NS 5 brane'' is ${\bf R}^4 \times
S^2$, where the two-sphere $S^2$ is wrapped on the $S^3$.  The
formation of the non-Abelian  configuration is energetically
favorable, because of the presence of the three-form flux; one may think of the branes, pointlike on the compact space, expanding into two-spheres under the influence of the flux background. This is the famous Myers effect \cite{Myers}.

We review how this works.  The $\overline{\rm D3}$-brane effective action (\ref{nbi3}) has the special property that in an imaginary $\it anti$ self-dual flux background, the
cubic terms in the full potential for the worldvolume fields $\Phi$ coming from the flux cancel. In our imaginary self-dual
flux background, on the other hand, there is no cancellation. Instead one finds \be \label{mypot}
V_{\rm ef\! f}(\Phi) \, \simeq\, \, {\mu_3 \over g_s} \Bigl( p -
i{4\pi^2 f \over 3} \epsilon_{ijk}
\Tr\Bigl( [ \Phi^k, \Phi^j] \Phi^l \Bigr) - {\pi^2 \over g_s^2}
\Tr\Bigl( [ \Phi^i, \Phi^j]^2 \Bigr) + \ldots\, \Bigr) \ .
\ee
As in \cite{Myers}, this potential has extrema away from the
origin $\Phi = 0$.  It is easy to verify that constant matrices
$\Phi^i$ satisfying the commutation relations \be \label{commreln}
[\Phi^i,\Phi^j] ~=~-{i g_s^2 f }\, \epsilon_{ijk} \Phi^k  \ee
represent a static solution to the equations of motion of
(\ref{mypot}). Up to rescaling, (\ref{commreln}) are just the
commutation relations which are satisfied by a $p\! \times\! p$-dimensional matrix representation of the $SU(2)$ generators
$[J^i,J^j] = 2i\epsilon_{ijk} J^k$. So by setting $\Phi^i =
-{1\over 2} {g_s^2 f} \, J^i$, with $J^i$ the generators of any
$p$-dimensional $SU(2)$ representation, we find a large class of
solutions of (\ref{commreln}).  Each $d$-dimensional irrep comprising the
$p$-dimensional representation should be thought of as a separate fuzzy sphere
composed of $d$ branes, and the location of the center of each is a flat direction.  Myers showed that the $p$-dimensional irreducible representation, where all the branes have coalesced, is the lowest-energy configuration.

The landscape of such fuzzy-sphere vacua is quite intricate, and
was analyzed in some detail in the work of Jatkar, Mandal, Wadia
and Yogendran \cite{JMWY}, who studied conditions under which reducible $SU(2)$
representations can roll perturbatively to the $p$-dimensional
irrep. JMWY found that when the fuzzy spheres are nested with the same center there is no tachyon, but when their centers are separated by a certain amount along the flat direction,
a path downward opens up.\footnote{This conclusion changes somewhat if a mass term $m^2 \Tr\Phi_i^2$ is added to the effective potential.
In this case, the flat directions are lifted and there is no classical path from nested reducible
reps to the irrep; however, such a path always exists for initially well-separated fuzzy spheres.}  It then follows that one can roll classically in the field space from the
configuration with $p$ separated anti-D3s, to the ``most giant''
NS5 which we wish to consider. It would be interesting to explore
whether inflation can occur in the convoluted route that one takes
through the fuzzy landscape of \cite{JMWY} to the final endpoint,
but we will not consider that question here.  We will instead
focus on the dynamics of the NS5-brane collective coordinate
$\psi$, which should capture the physics once the fuzzy sphere is
large enough.

To understand the motion of the non-Abelian inflaton when it is
still small, the gauge theory language is inadequate. Instead we
must use the dual supergravity description. The geometry
sufficiently close to a stack of $\overline{\rm D3}$-branes is a
Polchinski-Strassler-type throat, inside of which the stack
non-abelianizes into a giant inflaton 5-brane \cite{PS}. To
describe the evolution of the system and investigate its potential
use as an inflationary scenario, we must understand the
supergravity solution inside this throat region. We will study
this geometry and the resulting 5-brane potential in \S4.

\bigskip
\bigskip

\noindent {\sc Stage III: Rolling Giant Inflaton}

\begin{figure}
\centerline{\psfig{figure=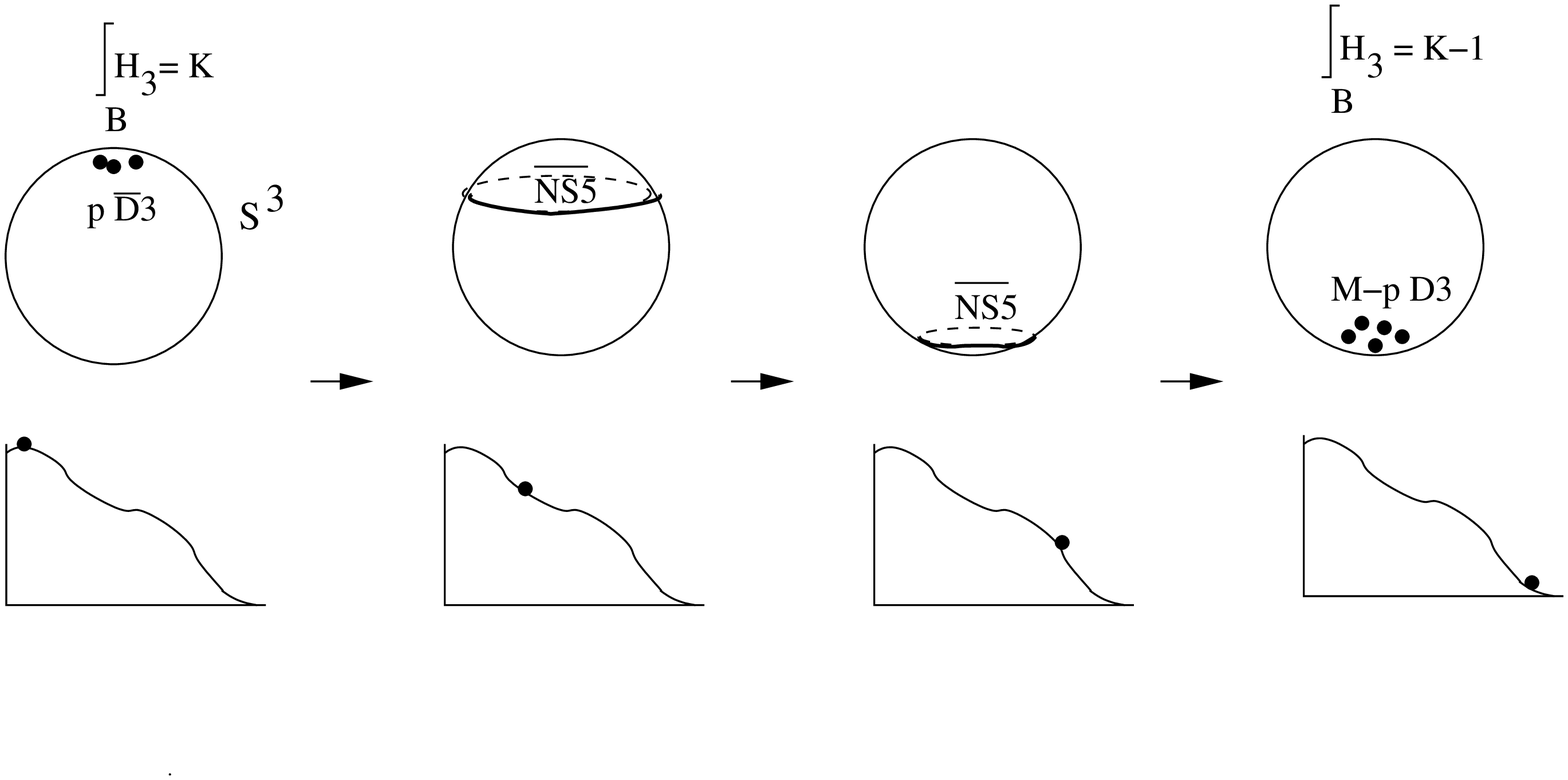,width=5.5in}}
\caption{The giant inflaton starts as a
bound state of $p$ anti-D3's, and expands due to the 3-form flux.
Near the slow roll region of the potential, its expansion slows
down due to a balance between the 5-brane tension and the
dielectric force. Eventually, the 5-brane decays to a
supersymmetric state with $M\!-\!p$ D3-branes.}
\label{s11Fig}
\end{figure}

We already reviewed the Myers effect by which the anti-D3s puff up
into a fuzzy 5-brane. As the size of the fuzzy $S^2$
grows, we expect a dual picture in terms of a wrapped NS5-brane
to become the most effective description of the system, as in
\cite{KPV}.
Let us parameterize the metric on the $S^3$ as
\be d \Omega_3^2 = d\psi^2
+ \sin^2 \psi \, d\Omega_2^2 \,.
\ee
We consider an NS5-brane, with anti-D3 charge $p$, wrapped around
the $S^2$ at the location $\psi$.  The anti-D3 charge is represented
by a flux of the worldvolume electro-magnetic field-strength $F=dA$
through the $S^2$.  The total potential for the motion of the 5-brane across the 3-sphere is \cite{KPV}
\ba
\label{effp} V_{\rm ef\!f} (\psi) = \, {\mu_3
 M \over g_s} \, \Bigl( \, {V}_2(\psi) + {1\over \pi} \, U(\psi)\Bigr) \,,
\ea
where we defined
\ba
\label{Potentials}
V_2 (\psi) \equiv {1 \over \pi} \sqrt{\sin^4 \psi + U(\psi)^2 }\,,
\quad & & \quad U(\psi) \equiv  {\pi p \over M} - \psi + {1 \over 2} \sin 2
\psi \,.
 \ea
\begin{figure}
\hspace{1.3in} \psfig{figure=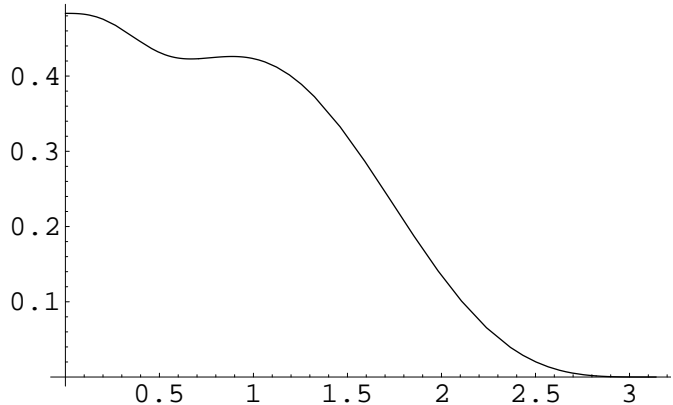,width=2.5in}
\caption{The effective potential $V_{\rm
ef\! f}(\psi)$ near the critical value for
 ${p\over M}\simeq 8 \%$, with only a marginally stable minimum. For smaller
 $p/M$ there is a more pronounced metastable vacuum, for larger
 $p/M$, the potential is monotonic.}
\label{roll2Fig}
\end{figure}
This potential is plotted in figure~\ref{roll2Fig}. The crucial property is that
for $p/M \lesssim .08$ it exhibits a metastable minimum, while for
$p/M \gtrsim .08$, the slope of the effective potential is
negative definite! In both cases we can draw an interesting
conclusion. In the regime with $p/M \lesssim .08$, the branes
reach a meta-stable state, corresponding to a static NS 5-brane
wrapping an $S^2$ of approximate radius $R^2 = g_s M
\sin^2\psi_{min}$. This state will eventually decay via quantum
mechanical tunneling to a supersymmetric state. In the regime ${p/
M} \gtrsim .08$, on the other hand, the nonsupersymmetric
configuration of $p$ ${\overline{D3}}$ branes relaxes to the
supersymmetric minimum via a {\it classical} process: the
anti-branes cluster to form the maximal size ``fuzzy'' NS 5-brane,
which then rolls down towards the bottom of the potential, at the
north-pole $\psi=\pi$. The end result of the process is $M\!-\!p$
D3-branes (in place of the original $p$ anti-D3-branes) while the
$H_{\it 3}$ flux around the B-cycle has been changed from $K$ to
$K\!-\!1$; it is hence referred to as brane/flux annihilation, and is depicted schematically in figure~\ref{s11Fig}.

This classical decay is our exit mechanism.  In addition, we see
that for $p/M$ very close to the critical value, the potential
exhibits an interesting plateau region near $\psi \simeq 0.7$.
Whether this region is sufficiently flat to support inflation
depends on the relative ratio of the string scale and the Planck
scale. In \S5 we determine the necessary bound on this ratio, and
in \S6 we discuss whether this bound can be satisfied.

The region of the NS5 potential (\ref{effp}) near $\psi=0$ also looks like a promising regime for
a slow roll.  As we have just
discussed, however, the NS5-brane description is expected to suffer large
corrections near $\psi=0$, because the gravitational backreaction
cannot be ignored.  Taking this backreaction into account is the goal of  \S4.

\newsection{Interbrane Attraction from Flux Polarization}

In this section, we will compute the leading order polarization of the background ISD three-form flux on the $S^3$ by a stack of $\overline{\rm D3}$-branes, and demonstrate how this induces an attractive force on other anti-branes.  We find it useful to define the following combinations of supergravity fields:
\be
\Phi_\pm \equiv e^{4A} \pm \alpha \,, \quad \quad G_\pm \equiv i G
\pm *_6 G \,.
\ee
The supergravity equations of motion then become (we assume $\tau = i/g_s$ for simplicity)
\begin{eqnarray}
\label{EOM1}
\tilde\nabla^2 \Phi_\pm = {g_s^2 e^{2A} \over 24} \,
|{G}_\pm|^2 + e^{-6A} |\nabla \Phi_\pm|^2
+ 4 g_s
\kappa_{10}^2 \mu_3 { e^{2A} \over \sqrt{g_6} } \sum_{i_\pm} \delta^6(y - y^{i_\pm}) \,,
\end{eqnarray}
\vskip-.35in
\begin{eqnarray}
d(\Phi_+ G_-)\; = \; d( \Phi_- G_+)_{} \,, \label{EOM2}
\end{eqnarray}
where $\kappa_{10}^2$ is the 10D gravitational constant and  $i_+$ and $i_-$ label D3- and $\overline{\rm D3}$-branes,
respectively.  The branes couple to the bulk fields as
\begin{eqnarray}
S_{3\pm} = - {T_{3} \over g_s} \int d^4x \sqrt{g_4} \, \Phi_\mp
\,.
\end{eqnarray}
We see that a $\overline{\rm D3}$ feels a potential from $\Phi_+$, while
it acts as a source for $\Phi_-$, and vice versa for a D3.

We are interested in evaluating the backreaction of the
anti-branes on the geometry near the apex. The unperturbed
background is imaginary self-dual, $\Phi^0_- = G^0_- = 0$.
Ignoring the anti-brane sources, this background trivially
satisfies two of the above equations, and the remaining equation
determines $\Phi_ + \equiv \Phi_+^0$ from a given $G_+ \equiv
G_+^0$. In our case, the resulting $\Phi_+^0$ is the warping of
the KS throat. We have $\Phi_+^0= 2 \anot^4$ at the tip and
$G_+^0$ as given in (\ref{flux1}).

Now let us include the effect of the anti-branes. It is clear that
they will immediately generate a $\Phi_-$ perturbation. This
perturbation, however, does not yet produce a force on the other
anti-branes. The question is whether, via coupling to the fluxes,
a change in $\Phi_+$ is induced as well.

We find it convenient to take advantage of the shift symmetry
$\alpha \rightarrow \alpha$ + const present in the equations of
motion.  Using this, we may shift $\Phi_+^0 \rightarrow 0$ at the
apex, while making $\Phi_-^0 = 2 \anot^4$. Since furthermore
$d\Phi_+^0 =0$ at the tip, we will ignore $\Phi^0_+$ in
calculating the leading perturbation induced to $\Phi_-$. For
$\Phi^0_+ = 0$, we may write the $\Phi_-$ equation as \be
\label{PhiMinusEqn} - \tilde\nabla^2 (\Phi_-)^{-1} = {g_s^2 \over
96} \widetilde{|G_-|}^2
+ 8 \pi^4 g_s {1
\over \sqrt{\tilde{g}_6} } \sum_{i_-} \delta^6(y - y^{i_-}) \,,
\ee
where the tilde indicates contraction with $\tilde{g}_{mn}$.
This form is very useful because all powers of the warp factor
have disappeared from the right-hand side.
Solving
(\ref{PhiMinusEqn}) in the presence of $p$ anti-branes ($G_-$ will
arise only as a perturbation and is subleading) we find
\begin{eqnarray}
\label{PhiMinusSoln}
\Phi_- = 2 \anot^4 \left({ y^4 \over y^4 + 4 \pi g_s p} \right)
\,, \end{eqnarray}
where an integration constant was chosen to give $\Phi_-^0 = 2 \anot^4$ for large $y$, and $y^2$ is defined with the warped metric.
This is nothing but the familiar geometry of a set of 3-branes in
flat space, approaching warp factor $\anot$ instead of $1$ far
away.  Thus the first effect of the anti-brane backreaction is to form
a new, small warped region deep inside the original geometry, as in \cite{HolComp};
this region can be viewed as a perturbation of the KS throat as long as $p \ll KM$.

The characteristic length scale of (\ref{PhiMinusSoln}) is $R_p^4 \equiv 4 \pi g_s
p$.   For $y^4 \gg R_p^4$, one is well outside this $\overline{\rm D3}$ throat region, and
one has
\begin{eqnarray}
\label{PhiMinusPert} \Phi_- \equiv 2 \anot^4 + \phi_- \simeq  2\anot^4 - {8 \pi g_s p
\anot^4 \over y^4}   \,,
\end{eqnarray}
where we defined the perturbation $\phi_-$.

The flux background will respond to the development of the
anti-brane throat.  In our conventions where $\Phi_+^0 = d \Phi_+^0
=0$ on the $S^3$, we have the leading order $G$
equation
\begin{eqnarray}
d (\Phi_- G_+) = 0 \,.
\end{eqnarray}
One then finds the solution for the three-form
\begin{eqnarray}
\label{GSoln} G_+ 
= \left( 1 + {4 \pi g_s p \over y^4} \right) G_+^0 \,.
\end{eqnarray}
The Bianchi identity $dG_+ = - dG_-$ requires $G_-$ to be turned
on as well. We will discuss the form of the $G_-$ flux in \S4.

Finally, the nonzero $G_+$ flux backreacts on $\Phi_+$, which we
have taken to vanish thus far, leading to a source in (\ref{EOM1})
proportional to
\be
|G_+|^2 \; \simeq \; |G_+^0|^2 \left( 1 - {2 \phi_- \over \Phi_+^0} 
\right) \; \simeq \; |G_+^0|^2 \left( 1 + {8 \pi g_s p \over y^4} 
\right) \,.
\ee
The leading piece already generated the KS throat, while
the subleading piece will produce a perturbation $\phi_+$ of
$\Phi_+$ via the equation of motion (\ref{EOM1}) \be
\tilde\nabla^2 \phi_+ = \; { \pi \anot^6 g_s^3 p \over 3 } \;
|G_+^0|^2\,  {1 \over \tilde{y}^4} \,. \ee Using $\nabla^2 (1/x^2)
= - 4/x^4$ and $|G_+^0|^2 = 24/(g_s^3 M)$, we  find \be
\label{PhiPlus} \phi_+  = - {2 \pi g_s p \over g_s M} \;
{\anot^4 \over y^2} \,. \ee Thanks to this perturbation, a test
anti-brane will indeed feel a force from the stack of $p$
$\overline{\rm D3}$s. This is the main result of this
section.

It is useful to compare (\ref{PhiPlus}) to the $\phi_+$ perturbation
that would have been created by a stack of $p$ D3-branes, instead of
anti-D3 branes; this is equal to the $\phi_-$ perturbation we found in
(\ref{PhiMinusPert}). The sign on the perturbations is the same,
so the force from the $\overline{\rm D3}$s is also attractive.
One can then define an effective D3-brane charge corresponding to
the $\phi_+$ perturbation (\ref{PhiPlus}),
\ba
\label{EffectiveCharge} Q_{D3} \, = \; {p\; y^2 \over 4 g_s M} \,.
\ea
This induced D3-brane charge results in an attractive force that is
weaker than the brane/anti-brane attraction at short distance,
but becomes comparable in magnitude at sufficient distance: recall that
$\!\sqrt{g_s M}$ is the characteristic length scale of the $S^3$, so this crossover happens on order the size of the available space.

\begin{figure}
\centerline{\psfig{figure=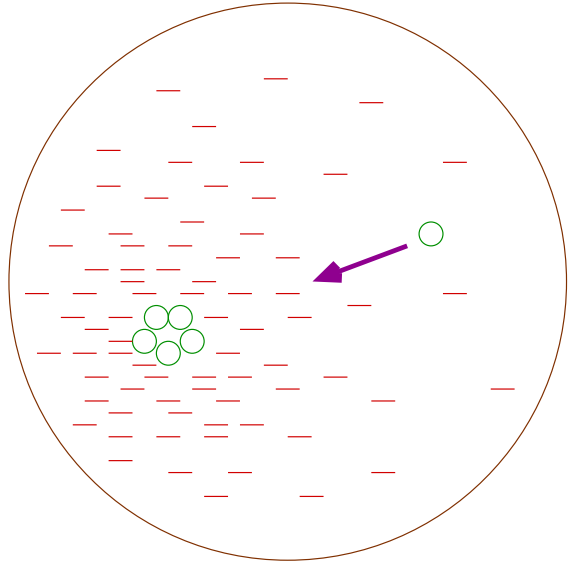}}
\caption{A stack of $\overline{\rm D3}$-branes polarizes the sea of flux, represented by dashes, leading to a force on a test $\overline{\rm D3}$-brane.}
\label{PolarizationFig}
\end{figure}

One may intuitively understand these results as follows. We can think
of the flux background effectively as a sea of D3-branes; it carries
D3 charge, as well as some energy-momentum.  When a stack of
$\overline{\rm D3}$-branes is placed in this sea, the background
adjusts itself in an attempt to screen the branes, by moving some of the
flux closer to the stack.  The effective charge of the
$\overline{\rm D3}$s is hence reduced, but the stress-energy in
their vicinity only becomes greater.  Consequently a test
$\overline{\rm D3}$-brane will feel a stronger gravitational
attraction than Ramond-Ramond repulsion, and will be drawn towards
the anti-branes.  (See figure~\ref{PolarizationFig}.) Because of the universal gravitational
attraction, the stack is never truly screened, and the effective
force only grows larger as more flux is displaced.  Moving further
away from the stack a greater volume of polarized flux is
enclosed, explaining the growth of the effective D3-brane charge
with distance (\ref{EffectiveCharge}).

\bigskip
\bigskip

\noindent
{\sc Condition for Slow Roll Inflation}

We have developed a physical picture: a test $\overline{\rm
D3}$-brane feels an attractive force from another $\overline{\rm
D3}$-brane due to the polarization of the background flux. The
force falls off with distance like $1/y^3$. We will now formulate
the condition for slow roll during the resulting motion of the
branes, which could last until the exit via nonabelianization is triggered, making this potentially a kind of hybrid inflation stage \cite{hybrid}.

Recall that the slow roll parameter $\eta$, which typically
imposes the most strict constraints on the potential, is defined
as \begin{eqnarray} \eta \, \equiv \, \MMp^2\; {V''\over V} \,. \end{eqnarray} where $V''$
denotes the second derivative of the inflaton potential, defined such that the inflaton kinetic term
is canonically normalized. We would like to apply this
prescription to our situation.

Starting from our initial condition with $p$ anti-branes scattered
randomly over the $S^3$, the process of forming a cluster goes in
successive steps. First the branes that are nearest to each other
form small clumps, which continue to merge with other small
clumps until the maximum size cluster is reached. An important
difference with the case of brane/anti-brane inflation is that each
small cluster retains its non-zero vacuum energy, and only
supercritical size clusters can decay and dump their vacuum energy
via brane/flux annihilation. How should we choose to parameterize the
inflaton field and compute the corresponding slow roll parameters
during the accumulation process?

A natural choice for the inflaton field $X$ is to take the square
root of the average (distance)$^2$ between the branes,
\begin{eqnarray}
 X^2 = {1\over p^2} \sum_{i\neq j} (y_i - y_j)^2.
 \end{eqnarray}
In the case that the branes
are uniformly distributed over the $S^3$, one has $X^2 = 2 R^2$,
where $R^2 = g_s M$ is the radius of the $S^3$. Given the
interbrane potential, which we denote ${\cal V}( y_i-y_j)$, it is possible to compute the
average static force on $X$. This computation is outlined in
Appendix B, with the following result
\be \label{supp} a_0^2
\ddot{X} \, \simeq \, -p\, {\cal V}'(X) \Bigl(1-{X^2 \over 2R^2}\Bigr)\,
\equiv \, a_0^2 V_{\rm eff}'(X) , \quad \quad R^2 = g_s M \ee where the interbrane potential
${\cal V}(X)$ reads \be
{\cal V}(X)\, = \, {2\pi g_s \over g_s M} {\anot^4\over X^2} \,.\ee This result
has the expected feature that for a uniform brane distribution, so that $X^2 = 2 R^2$, the force vanishes. We now compute
$V''(X)$ by differentiating $V_{\rm eff}'(X)$ at $X^2 = 2R^2$. We
obtain \be \label{DistribMass} V''(X) \simeq {2 \pi p g_s \anot^2 \over (g_s M)^3}\, .
\ee The total potential $V$ is twice the energy stored in the
anti-brane tension, $V = 2 p T_3 \anot^4 / g_s$. Putting things
together we find \be \label{eta1}
|\eta| \, \simeq\; {16\pi^4 \over
g_s M^3}\; {\MMp^2 \over \anot^2 \MMs^2}\,,  \ee 
where we have restored $M_s$, previously set equal to 1; $a_0 M_s$ is the string scale at the bottom of the throat.  So we would get the
required amount of inflation in case we could embed our scenario
in a rather mildly warped setting, such that \be \label{ineq}
{\anot^2 \MMs^2 \over \MMp^2} \, \gtrsim {10^5 \over g_s M^3} \,. \ee
As mentioned earlier, $M$ can be chosen as large as $10^3$ (or
even larger). Taking $g_s \sim 10^{-1}$, we find that $a_0 \MMs$
can be a factor of $30$ below the Planck scale. As we will discuss
in more detail in \S6, this is difficult to realize within the regime of validity of our
approximations.

\newsection{Gravity Dual of the Non-Abelian Inflaton}

The preceding analysis is only valid so long as the backreaction is small, which is the case outside the anti-brane throat, $y^4
\gg 4 \pi g_s p$. As one goes down the throat, $\Phi_-$, which was
growing as we approached the throat,  ``turns around" and begins
decreasing as (see (\ref{PhiMinusSoln}))
\begin{eqnarray}
\Phi_- \simeq  {2 \anot^4 y^4 \over 4 \pi g_s p} \,,
\end{eqnarray}
which is the usual result for the near-horizon geometry of the
stack of branes. The three-form flux (\ref{GSoln}), however, is forced into
blowing up as $G_+ \sim 1/y^4$ to compensate for $\Phi_-$ in the
$G$ equations of motion (\ref{EOM2}). We see that once we are within the throat,
the fluxes are no longer small and our approximations of the last section break down.  What can we learn about the
geometry near the anti-branes?

The perturbation of the near-horizon throat of a stack of
D3-branes by 3-form flux has been studied in the classic paper of
Polchinski and Strassler (PS) \cite{PS}.  PS
found (see sec.~III.D) four linearized solutions for $G_3$,
falling off as powers $y^p$ with $p = 0,-4,-6,-10$.  The $p=0,-10$
solutions are associated with a constant IASD tensor\footnote{Here and in the following we have exchanged $G_+$ and $G_-$ in the PS solutions to adapt them to our case of an
anti-brane (rather than a brane) throat.}, which is not our
situation.
The $p=-4,-6$ solutions, on the other hand,  are
constructed from a constant ISD tensor $T_3$ in the anti-brane
throat.  The solutions are, for $p=-4$,
\begin{eqnarray}
\label{4Soln} G_+ = {- i 2\sqrt{2} \over g_s}{R^4 \over y^4} T_3
\,, \quad \quad G_- = {-i 4\sqrt{2}  \over g_s}{ R^4 \over y^4}
(T_3 - 2 V_3) \,,
\end{eqnarray}
and for $p=-6$,
\begin{eqnarray}
\label{6Soln} G_+ = 0 \,, \quad \quad G_- = \varsigma {R^6 \over
y^6} (T_3 - 2 V_3) \,,
\end{eqnarray}
where $V_{mnp} \equiv {y^q \over y^2} ( y^m T_{qnp} + y^n T_{mqp}
+ y^p T_{mnq})$; one may check that $T_3 - 2 V_3$ is indeed IASD.

We see that our leading $G$ perturbation (\ref{GSoln}), which we
found by matching to the region outside the throat, is of the
$p=-4$ form (\ref{4Soln}), with ISD tensor
\be
T_{ijk} \simeq {\epsilon_{ijk}  \over \sqrt{2 g_s M}} \,.
\ee
This $p=-4$ solution corresponds in the holographic dual to the
addition of a quadratic term in the superpotential of the non-Abelian
worldvolume gauge theory, which generates
a cubic term in the full potential, of the same form as the matrix
potential given in eqn (\ref{mypot}). In addition, it generates masses
for the fermions and bosons, proportional to $1/\sqrt{2g_s M}$.
The worldvolume gauge theory description has limited validity, however,
since it is strongly coupled. Instead, the system must be studied
using the dual supergravity.

Polchinski and Strassler solved for the effect of the flux perturbation
on the supergravity geometry and on the location of the $p$ branes
generating the throat.  Their essential result is that the branes tend
to become non-Abelian and balloon up into an 5-brane wrapping a
transverse $S^2$. The radial motion of the 5-brane is governed
by a effective potential drawing it to a certain minimal energy
location within the throat geometry. It was further shown that, due to some miraculous cancellations, the exact
form of the effective potential is reproduced by a simple probe
calculation based on a single 5-brane moving in the original
throat geometry.

The result that the branes tend to non-abelianize
is, of course, consistent with our own physical picture.
The additional lesson that we have now learned, however, is that
the potential obtained in \cite{PS} is the proper refinement of
the 5-brane potential $V_{\rm ef\!f}(\psi)$ given in (\ref{effp})
in the region near $\psi = 0$, where the backreaction needs to
be taken into account.

This 5-brane potential can be found in section IV.C of \cite{PS},
eqn.~(72).  We match their $z$ as $|z|^2 = y^2/2$, and for the
NS5-brane, $z$ is real.\footnote{This $z$ is a complex coordinate
in \cite{PS} and should not be confused with the complex structure
modulus introduced in \S6.} The potential then becomes \be \label{PSPotential} V(y)
= \alpha y^2 (y - \beta)^2 \,, \quad \quad \alpha =  {\anot^4
\over 16 \pi^5 g_s^3 p} \,, \quad \quad \beta = {\pi g_s p\over
\sqrt{2 g_s M}} \,. \ee The quadratic term in (\ref{PSPotential})
is fixed in \cite{PS} by supersymmetry. In our situation,
supersymmetry is broken by the conflict between the anti-brane
throat and the surrounding ISD background.  One may wonder,
therefore, whether the quadratic term will be absent or modified
in our case. This term has a direct dynamical origin, however, in
the backreaction of the fluxes on $\Phi_+$. We should expect to
obtain the same result (\ref{PSPotential}) in any regime where our
flux and geometry agree with that of \cite{PS}.

The potential (\ref{PSPotential}) has two minima,
at $y =0$ and $y = \beta$.  The former is outside the validity of
the supergravity approximation, while the latter is
the location where the giant comes to sit. The radius for the
giant is
\be
\label{radius}
y_0^2 = {\pi^2 g_s p^2 \over 2  M} \,,
\ee
This answer should be compared with the estimate
in \cite{KPV}, eqn.~(32): $y^2 = 4 \pi^2 g_s (p^2~-~1)/M$,
which was obtained from the non-Abelian theory, neglecting the
gravitational backreaction. One sees that the parametric dependence
matches nicely; the difference in the constants can be interpreted as the tendency of the 5-brane to be held back by its own backreaction. Both results, however, apply only in the limit
where $p\ll M$. It is easy to verify from the shape of the potential (\ref{effp}), that in case $p/M$ gets
close to the critical value $p_{crit}/M \sim .08$, the size of
the giant graviton in fact starts to exceed its gravitational radius.
Since this is the regime we are interested in, we must conclude
that for the near-critical value of $p/M$, the PS potential
(\ref{PSPotential}) can be trusted only for $y$ sufficiently smaller
than $y_0$ given in (\ref{radius}).

Meanwhile, (\ref{PSPotential}) also exhibits a maximum at $y=
\beta/2$. It can be shown that this maximum occurs within the
regime of validity of the supergravity approximation, and is also
just far enough down the throat, so that the PS potential provides
a good description. We would like to investigate whether this top
of the potential is a viable starting point for a slow-roll
evolution of the giant inflaton 5-brane. Can we get a small value
for the inflationary parameter $\eta$ there?

Our coordinate $y$ is not canonically normalized:  the $y$
kinetic term is proportional to
\begin{eqnarray}
S_{kin} = - {T_3 p \, \anot^2 \over g_s} \int\! d^4x \,
{1 \over 2}(\partial_\mu y)^2 \,.
\end{eqnarray}
At the maximum $y=\beta/2$, we find using (\ref{PSPotential})
\be
{\partial^2V\over \partial y^2} \, \simeq \,
{p \, \anot^4 \over 32\pi^3 M g^2_s}\qquad  \rightarrow \qquad V''
\, \simeq\; {\anot^2\over 4g_s M} \,.
 \ee
To estimate the value of the potential $V$ at this maximum, note
the contribution $V(\beta/2)$ from the PS potential is much
smaller than the overall contribution from the anti-brane tension,
$V = 2 p T_3 \anot^4 / g_s$. Using these facts, a straightforward
calculation gives \be \label{realeta} |\eta|\, \simeq \,\, {\pi^3
\over p M}\; {\MMp^2 \over \anot^2 \MMs^2} \ee Inflation works
provided $|\eta|\, \lesssim \, 1/30$, which requires that the ratio
of the red-shifted string scale $\anot \MMs$ at the bottom of the
conifold and the 4-d Planck scale must satisfy the inequality \be
\label{condition1} {a^2_0 \MMs^2 \over \MMp^2} \; \gtrsim \; {10^3
\over p\, M} \, . \ee A possible value of $p \, M$ is of order
$10^5$. In this case, this inequality implies that $\anot
\MMs$ is just one order of magnitude below the 4-d Planck scale.

\newsection{Numerical Study of the Rolling Giant Inflaton}

Up to now our focus has been on the
dynamics at the onset of NS5 brane formation.
We now wish to consider the possibility of inflation produced during the rolling
phase of the giant inflaton. Examining the
potential (\ref{effp}) indeed reveals another promising regime (well
studied with the NS5 action). For very small $p/M$, there is a
metastable giant graviton vacuum at finite $\psi$. As one
increases $p$, there is a critical value $p_{crit}$ above which
the metastable vacuum disappears -- the anti-D3 branes
perturbatively roll to $M-p$ D3 branes, a feature which provides
the graceful exit of our inflationary model. As a
consequence of this structure, for $p \sim p_{crit}$ there is
actually a plateau in the potential (\ref{effp}) at intermediate
values of $\psi$.  This plateau can be used to provide several
e-foldings of inflation at intermediate $\psi$.   Hence, the system
of $p$ anti-D3s in the warped flux background is rich enough to
potentially exhibit several inflationary phases.

Because the dynamics are more involved in the plateau region,
we will study them by explicitly setting up the coupled system of
scalar and Friedmann equations, and solving
these numerically using Mathematica.  We find that for fixed $p/M$, the
physics is controlled by only one nontrivial parameter (which we call $B$).
For clarity, we now derive the explicit form of the equations of motion
that we used for numerical integration.

The 5-brane world-volume action reads  \cite{KPV}
\begin{eqnarray}
\label{NS5Action} S_{NS5} = - \ANOT \int d^4x \sqrt{-g_4} \left[
V_2(\psi) \sqrt{1 - Z^2 \dot\psi^2} + {1 \over \pi} U(\psi)
\right] \,,
\end{eqnarray}
with $V_2(\psi)$ and $U(\psi)$ as in (\ref{Potentials}), and
\ba
 \ANOT \equiv {\mu_3 M \anot^4 \over g_s} \,, \quad \quad Z^2
\equiv {g_s M \over \anot^2} \,. \ea
The 5-brane equations of motion are most conveniently expressed in first-order
Hamiltonian form. The conjugate momentum derived from (\ref{NS5Action}) is
\begin{eqnarray}
P = \ANOT V_2(\psi) { Z^2 \dot\psi \over \sqrt{1 - Z^2
\dot\psi^2}} \,,
\end{eqnarray}
leading to the Hamiltonian \be {\cal H}= \sqrt{P^2/Z^2 +
\ANOT^2 \, V_2(\psi)^2} + {\ANOT \over \pi} U(\psi) \,. \ee
Hamilton's equations are \ba \label{PositionHamilton} \dot\psi =
{\partial{\cal H} \over
\partial P} &=&
{P \over Z^2 \sqrt{ P^2/Z^2 + \ANOT^2 \, V_2(\psi)^2}} \,, \\[2.5mm]
\label{MomentumHamilton} \dot{P} = - {\partial {\cal H} \over
\partial \psi} &=& {\ANOT \over \pi}(\cos 2\psi - 1) - {\ANOT^2 \, (4
\sin^3 \psi \cos \psi + 2 U(\psi) (\cos 2 \psi -1)) \over 2 \pi^2
\sqrt{P^2/Z^2 + \ANOT^2 \, V_2(\psi)^2}} \,. \nonumber \ea We
couple to 4-d gravity: \be S_{tot} = \MMp^2 \int d^4x \sqrt{-g_4}
{\cal R} + S_{NS5} \,, \ee and assume a flat FRW universe, \be g_{\mu\nu} dx^\mu dx^\nu = -dt^2 + a(t)^2 d\vec{x}^2 \,. \ee Since
we are assuming only time derivatives in $S_{NS5}$, the scale factor is present only in the overall
$\sqrt{-g_4}$.  Hence it can be taken into account by scaling
$\ANOT \rightarrow a^3 \ANOT$ in (\ref{PositionHamilton}),
(\ref{MomentumHamilton}).  The Friedmann equation is (written in
terms of the momentum $P$) \be \left({\dot{a} \over a}\right)^2 =
{ \mu_3 M \anot^4 \over 6 \pi g_s \MMp^2} \left[ \sqrt{P^2 \pi^2
/(\ANOT^3 a^6) + \sin^4 \psi + U(\psi)^2} + U(\psi) \right] \,.
\ee We find it convenient to define the variables $\tilde{a}^3
\equiv \ANOT a^3 /\pi$, $\tilde{P} \equiv P/Z$, in which case we
can write the three coupled first-order equations as \ba Z
\dot\psi &=& {\tilde{P} \over \sqrt{ \tilde{P}^2 + \tilde{a}^6 \,
\pi^2 V_2(\psi)^2}} \,, \nonumber
\\[2.5mm]
Z \dot{\tilde{P}} &=& \tilde{a}^3 (\cos 2\psi - 1) - {\tilde{a}^6
\, (4 \sin^3 \psi \cos \psi +
2 U(\psi) (\cos 2 \psi -1)) \over 2 \sqrt{\tilde{P}^2 + \tilde{a}^6 \, \pi^2 V_2(\psi)^2}} \,, \\[2.5mm]
Z \dot{\tilde{a}} &=& \tilde{a} \sqrt{B} \left[ \sqrt{\tilde{P}^2
\tilde{a}^{-6} + \sin^4 \psi + U(\psi)^2} + U(\psi)
\right]^{1/2} \,, \nonumber \ea with \be \label{BEqn} B \equiv {Z^2 \mu_3 M \anot^4
\over 6 \pi g_s \MMp^2} =
{M^2 \over 4 8 \pi^4} \; 
{\anot^2 \MMs^2 \over \MMp^2} \,.
\ee
We solved these equations numerically.
Notice that $Z$ now appears in combination with the time derivative, and thus
can be absorbed into a new definition of time.  Hence the relevant
parameters for controlling the dynamics are just $B$ and $Y \equiv
\pi p/M$; for $p = p_{crit}$ we have $Y \sim 2/7$.

\begin{figure}
\centerline{\psfig{figure=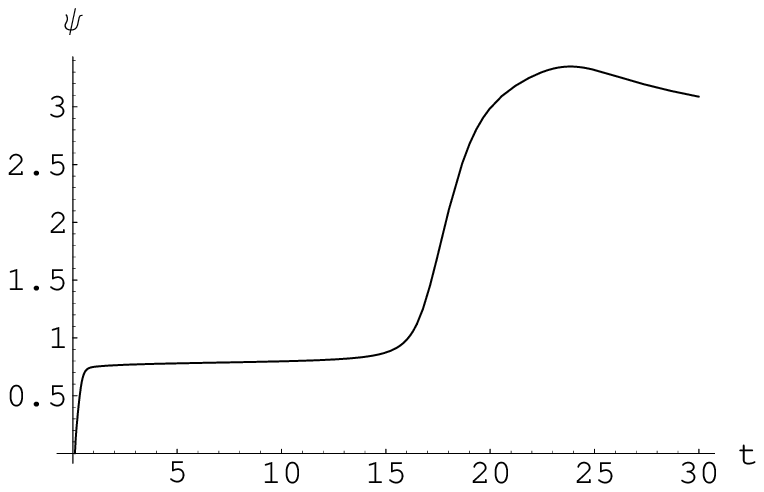,width=3in}\psfig{figure=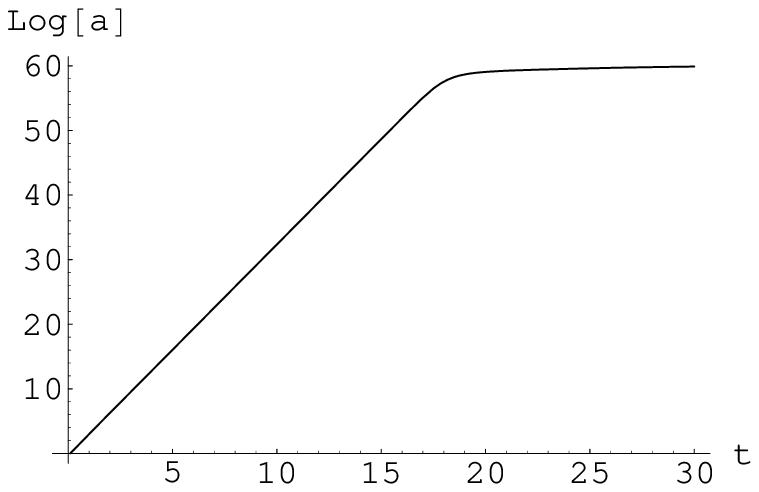,width=3in}}
\caption{The giant inflaton trajectory for $B=0.85$ and
initial conditions $\psi(0) = .03$ and $\tilde{P}(0)=0.3$, and the
corresponding evolution in the log of the scale factor $a$. We see
that almost all of the inflation comes from the shoulder region
near $\psi=0.7$.}
\label{ShoulderFig}
\end{figure}
In figure~\ref{ShoulderFig}, we have indicated a typical trajectory for $B \sim .85$,
and $Y=2/7$. The initial conditions chosen are $\psi(0)= 1/25$ and
$\tilde{P}(0)=0.3$. The evolution is insensitive to the initial condition of $\psi$ as long as it is near zero.   We see that one quite easily obtains $\sim 60$
e-foldings of exponential expansion, with quite generic initial
conditions. All of the expansion is generated in the shoulder
region, where the potential flattens out. If one allows for
smaller initial momenta, we find that one can still get around
60 e-foldings for values of $B \simeq 0.5$.

Hence given the expression (\ref{BEqn}) for $B$, we conclude that
the rolling giant inflaton can represent an interesting scenario
provided that it can be realized with a mild enough warp factor
$\anot$. The condition on $\anot$ is roughly \be
\label{condition2} {\anot \MMs \over \MMp} \gtrsim {65 \over M}\,
. \ee Note that this condition is slightly less stringent than
(\ref{condition1}), given that $p/M \sim 10^{-1}$. In the next
section we will analyze whether this condition can be satisfied
within our set-up.

\newsection{The Viability of the Giant Inflaton}

In the previous sections, we have expressed the conditions for
inflation in terms of specific inequalities (\ref{ineq}),
(\ref{condition1}) and (\ref{condition2}) for the ratio of the
red-shifted string scale $\anot \MMs$ and the 4-d Planck scale
$\MMp$. The inequalities also involve the microscopic parameters
$M$, $p$ and $g_s$; the ratio $a_0 \MMs / \MMp$ is not independent
of these quantities. In this final section we will study whether
the inequality can be satisfied within our set-up.

The 4-d Planck scale is expressed in string units as \be \MMp^2 =
{V_6\over g_s^2 \kappa_{10}^2} \,, \quad \quad \kappa_{10}^2 = \pi
(2\pi)^6  \,,\ee where $V_6$ is the warped volume of the
compactification manifold. We wish to obtain an estimate of the
minimal possible value of $V_6$ for given flux $M$ and $K$. To
this end, let us compute the warped volume of the throat region.
In the warped region between the tip and the Calabi-Yau manifold,
the throat geometry takes the approximate form \be
\label{ThroatGeom} \tilde{g}_{mn} dy^m dy^n = dy^2 + y^2
ds_{T^{1,1}}^2 \,, \quad e^{-4A} \approx {R^4 \over y^4} \,, \quad
R^4 \approx {27 \pi \over 4}\,  g_s M K \,, \ee giving a total
space that is approximately $AdS_5 \times T^{1,1}$, where
$T^{1,1}$ is the base of the conifold. We can now perform the
integral \ba
V_6 \equiv \int
\!\! d^6y\,\sqrt{\tilde{g}_6}\, e^{-4A} \! &\!
 \simeq \! & \! \int_{T^{1,1}}\! \!\! d\Omega\; \int_{y_0}^{y_1}\!\!\!\! dy \, y^5\,
 {R^4\over y^4} 
\ \simeq \ {1\over 2}\, {\rm vol}(T^{1,1}) \, R^6 \,,
\ea
where in the last line we assume that the location $y_0$ of the
bottom of the throat is small compared to the location $y_1 \sim R$
where the throat is capped off by the CY geometry. Plugging in the
values for $R^6$ and the known volume of $T^{1,1}$, we thus
obtain a lower bound for the total warped 6-volume, given by
\be
V_6 \gtrsim \Bigl({2\pi\over 3}\Bigr)^3 \Bigl({27\pi\over 4} \, g_s M K\Bigr)^{3/2}
\ee
The warp factor at the bottom of the KS throat scales with powers of the overall volume, as well as the complex structure $z$ of the conifold geometry, which is also determined by the microscopic parameters:
\be \anot^2 \simeq {V_6^{1/3} z^{2/3} \over g_s M} \;\,, \quad \quad
z \simeq e^{-{2\pi K \over g_s M}} \,. \ee
Combined we derive the following inequality for the ratio of the
warped string scale and the 4-d Planck scale: \be \label{upper}
{a_0^2 \MMs^2\over \MMp^2} \, = \, {g_s^2 \, \anot^2\,
\kappa_{10}^2\over V_6} \; \lesssim\; {64 \pi^4 z^{2/3} \over 3
M^2 K} \,. \ee This inequality should be compared with our conditions
(\ref{ineq}), (\ref{condition1}) (\ref{condition2}) for inflation.

The most promising stage for inflation, it turns out,  is stage I,
the accumulation process of the anti-branes on the $S^3$.
Combining the result (\ref{eta1}) of \S3 and the estimate
(\ref{upper}) we obtain the lower bound for $|\eta|$
during this stage \be \label{workre} |\eta| \gtrsim {{3 K \over 4g_s M}\,
\exp\Bigl({4\pi K \over 3 g_s M}\Bigr)}  \,.\ee Given this formula,
slow roll would require that $g_s M$ is at least 30 times larger
than $K$, which would correspond to a very shallow, mildly warped
throat.

Such a shallow throat is problematic for our approximations,
however.  Our description in terms of the conifold geometry holds only for $z
\ll 1$, which requires $K$ to be larger than $g_s M$.
This renders our conclusion that inflation works in the regime
(\ref{workre}) suspect.
For this reason, we will not try to analyze the inflationary predictions
in any detail.  It would be interesting (though technically challenging)
to study this scenario in a global setting where the 
calculations could be continued beyond our present regime of control.

The analogous results for the giant inflaton moving in its own throat (\ref{eta1}), and rolling over the shoulder (\ref{BEqn}) are
\begin{eqnarray}
\label{IIeta}|\eta| \simeq {3 M K \over 64 \pi p} \exp\Bigl({4\pi K \over 3 g_s M}\Bigr) \,,\end{eqnarray}
and\begin{eqnarray}
\label{IIIB}B^{-1} = {9 K \over 4} \exp\Bigl({4\pi K \over 3 g_s M}\Bigr) \,.\end{eqnarray}The throat roll result (\ref{IIeta}) is moderately larger than (\ref{workre}), by a factor $(g_s M)(M/p)/(16 \pi)$.  $B^{-1}$ is larger than (\ref{workre}) by $3 g_s M$, but as we discussed $B^{-1}$ can be as large as $1$ or $2$.  All three results are tantalizingly close to realizability, but lie just outside the bounds of our approximations.  It is intriguing to speculate that if we could gain control of the region $K \ll g_s M$, these giant inflaton scenarios could be realized.

\newsection{Discussion}

Our results illustrate several simple points about brane cosmology
in string theory.  Among them:

\noindent $\bullet$ Unlike the models described in
\cite{KKLMNOPQRST}, in the promising regime of parameters these
models provide inflation at a very high scale.  This exacerbates
the challenges of moduli stabilization (one must make sure the the
radion and dilaton are stiff already at this very high $V^{1/4}$),
but relaxes the tuning associated with obtaining initial
conditions appropriate for low-scale inflation.  Indeed, even if
the 60+ e-foldings which explain our flat, homogeneous bubble
occur at $V^{1/4} << \MMs$, it is natural to postulate a
primordial phase of inflation very close to $V^{1/4} \sim \MMs$ to
explain the initial conditions for the later stage.  The giant
inflaton could provide a model of this ``primordial inflation,''
which can occur at a very high scale and need not last for 60
e-foldings.

\noindent $\bullet$ Unlike the models described in
\cite{KKLMNOPQRST}, here we see that warping actually works $\it
against$ the success of many potential models.  We have always
assumed very mild warping $\anot$ in the inflationary throat,
because our slow-roll parameters scale like $1/\anot^2$.  The
reason for the difference between this class of models and the
models of \cite{KKLMNOPQRST} arises because only potentials which
are inverse power laws in the canonical inflaton field provide
improved inflationary properties in warped backgrounds. More
conventional field theoretic models with positive power-law
potentials (analogous to the effective field theories which arise
in our models) are hindered by the warping.

\noindent $\bullet$ The emergence of the standard model at the end
of inflation may be more ``stringy'' than is assumed in the most
conventional models.  For instance, in brane/anti-brane scenarios,
it is natural to assume that the standard model branes (plus an
extra) are the targets which collide with an anti-brane to end
inflation; then the standard-model open strings are naturally
excited in a reheating process during the brane/anti-brane
annihilation.  In such a model, the standard model degrees of
freedom are already evident as perturbative quantum fields during
inflation.  In a scenario like ours, it is possible for the
standard model to emerge on the $M-p$ D3 branes which only exist
$\it after$ the brane/flux annihilation is completed.  In this
sense the standard model degrees of freedom may only emerge as
perturbative objects at the end of inflation.

\noindent $\bullet$ One merit of exhibiting an inflationary
scenario within a microscopically complete theory is that it
allows one to examine the nature and severity of the required
tunings to produce inflation.  As in \cite{KKLMNOPQRST}, we find
that tuning is required to produce a working model in our
scenario.  However, the tuning can be explicitly parameterized in
terms of microscopic data which is at our disposal -- the choice
of the number of anti-D3 branes $p$ and the background RR
flux $M$.  The severity (or lack thereof) of the tuning may be
best estimated not by assuming a flat measure on e.g. $\eta$
space, but instead by asking: how severely must we tune the
microscopic parameters within their reasonable ranges, to obtain a
desirable value of $\eta$? In the case at hand, it looks a bit
worse than one would have expected, but there
is no reason to expect that this is a general feature.

\bigskip
\bigskip

\begin{center}
{\bf Acknowledgments}
\end{center}
\medskip
We would like to thank Kristin Burgess, Renata Kallosh, Andrei
Linde, John McGreevy, Eva Silverstein and Lenny Susskind for
discussions. We are particularly grateful to John Pearson for early collaboration.
This material is based upon work supported by the
National Science Foundation under grants No.~0243680 (O.D.
and H.V.) and PHY-0097915 (S.K.). Any opinions, findings, and
conclusions or recommendations expressed in  this material are
those of the authors and do not necessarily reflect the views of
the National Science Foundation. The work of S.K. was also
supported by a David and Lucile Packard Foundation Fellowship for
Science and Engineering, and the DOE under contract
DE-AC03-76SF00515.

\appendix
\sect{Forces generated by distant fluxes}

In compactifying the KS throat as in \cite{GKP}, one introduces
other fluxes elsewhere in the Calabi-Yau manifold.  These will
generically backreact and produce perturbations to the KS throat.
Here we show that these corrections break the Goldstone-mode shift
symmetry on the $S^3$ for the D-brane collective coordinates.  We
present two arguments: a direct gravity-side estimate, and a dual
holographic field theory estimate. The two agree.  The gravity
estimate basically uses the same logic as \cite{Uranga}, which
studied soft-breaking terms in flux compactifications.

\subsection{Anti-D3 potential from distant fluxes}

We consider ``distant fluxes" supported on cycles not associated with our throat, but
preserving the ISD property. The effect on the warp factor can be
determined from the equation of motion (\ref{EinsteinEq}) with
$\alpha = e^{4A}$ and ${\cal R}_4 = 0$,
\begin{eqnarray}
\label{WarpingFromFlux} \nabla^2 A = {g_s^2|G|^2 \over 48} \,,
\end{eqnarray}
where to this order we are ignoring the tension of the anti-branes.
The leading contribution to $|G|^2$ comes from the primary fluxes
$M$ and $K$, which of course generates the radial warp factor
(\ref{ThroatGeom}) respecting the $SO(4)$-symmetry. We consider the subleading
corrections involving the distant fluxes.

The primary flux at the base of the throat is equal to \be G_{mnp}
= {2 M \epsilon_{mnp} \over (g_s M)^{3/2}} \,, \ee in terms of
the warped epsilon tensor of the 3-sphere.
A natural estimate for the distant flux is that is proportional,
up to some factor $f$ of order unity, to the unwarped volume of
the $S^3$:
\be \delta G_{mnp} \sim f \Omega_{mnp}  \sim {f z
\epsilon_{mnp}\over (g_s M)^{3/2}} \,, \ee where we used that
$\int_A \Omega = z$. One way of thinking about this is that the density of
primary flux must be very large in unwarped units, since it is
integrated over a small cycle to obtain a fixed value $M$.  The
distant fluxes will generically be associated to a cycle of order
one, and hence the density of the flux will be smaller, by an
order $z$.

The subleading value of $|G|^2$ is then \be \label{DistantFlux}
{g_s^2 (G_{mnp} \, \delta\overline{G}^{mnp} + \delta G_{mnp}
\overline{G}^{mnp} ) \over 48}
\sim {f z \over 2 g_s M^2} \, \equiv \, {1\over 8} m_{{}_{\!
X}}^2 \anot^{-2}  \quad \rightarrow \quad m_X^2 = {4 f z \anot^2
\over g_s M^2} \,. \ee Considering only the variation of the
warp factor over the $S^3$, we then estimate \be A(\Phi) \sim
\ANOT + {1\over 8} m_{{}_{\! X}}^2 \anot^{-2} g_{ij} \Phi^i \Phi^j
+ \ldots \,, \ee and find that $m_X^2$ is the effective mass for
canonically normalized fields $X^i \equiv \anot \sqrt{T_3 M}
\Phi^i$,
\ba \label{NormalizedAction} S_{\overline{\rm D3}}& \sim & - \int
d^4x \sqrt{-g_4} \, \left( {2 T_3 p \anot^4 \over g_s} + {1\over
2} m_{{}_{\! X}}^2 \, {\rm Tr} \, X_i^2 + {1 \over 2}\, {\rm Tr}
\, (\partial_\mu X_i)^2\right) \,. \ea All mass-scales in the
above formulas are expressed in units of the unwarped string scale
$\MMs$. We see that beyond the overall redshift of $\anot$ that
affects all masses at the bottom of the throat, the mass-squared
$m_{{}_{\! X}}^2$ induced by distant fluxes is suppressed by an
additional factor of $z = \exp(-2 \pi K / 3 g_s M)$.

One may easily impose a discrete symmetry on the geometry such
that the crossterm (\ref{DistantFlux}) vanishes.  In this case,
the leading mass
correction is instead \be 
\label{O10DistantFlux} {g_s^2 \delta G_{mnp} \delta
\overline{G}^{mnp} \over 48} \sim { f^2 z^2 \over 8 g_s M^3}
\equiv {1 \over 8} m_X^2 \anot^{-2} \quad \rightarrow \quad m^2_X
=
{ f^2 z^2 \anot^2 \over g_s M^3} \,. \ee 
which is suppressed by  two factors of $z$.  The masses in (\ref{DistantFlux}), (\ref{O10DistantFlux}) are smaller than the effective mass from interbrane forces (\ref{DistribMass}), and hence we neglect them in our estimates of inflation.

\subsection{Holographic argument}

It is instructive to consider these symmetry breaking
perturbations from the point of view of the holographic dual
picture.  The Klebanov-Strassler geometry has a dual description
as a four-dimensional ${\cal N}=1$ $SU(M(K+1)) \times SU(MK)$
field theory with bi-fundamental fields $A_i$, $B_j$ transforming
in the $({\bf 2}, {\bf 1})$ and $({\bf 1}, {\bf 2})$ of $SO(4)
\sim SU(2) \times SU(2)$ and a quartic superpotential.  The rest
of the geometry at the top of the KS throat can be interpreted as
a ``Planck brane" in the spirit of \cite{RS}, corresponding to
additional dynamics cutting the theory off in the UV, at the
Planck scale.  This is realized as irrelevant operators suppressed
by powers of $\MMp$ added to the dual field theory.

The $SO(4)$-breaking physics of the distant fluxes is hence
translated into $SO(4)$-breaking irrelevant operators in the dual.
One can estimate these as follows. Assuming unbroken
supersymmetry, we consider corrections to the superpotential.  The
most straightforward class of these is (see \cite{KW}): \be \Delta
W_n = {\cal C}^{i_1 i_2 \ldots i_n j_i j_2 \ldots j_n}  \, {\rm
Tr} \left( A_{i_1} B_{j_1} A_{i_2} B_{j_2} \ldots A_{i_n} B_{j_n}
\right) \,. \ee 
For generic choices of ${\cal C}$, $SO(4)$ is broken.  Due to the
anomalous dimensions of the $A_i, B_j$ fields, these perturbations
have dimension $\Delta_n = 3n/2$. The superpotential of the
theory, which is marginal, is a special case of $n=2$.  Hence the
leading irrelevant operator has $n=3$ and dimension $\Delta =
9/2$, while the subleading irrelevant perturbation is $n=4$ with
$\Delta = 6$.  The corresponding terms in the component Lagrangian
have dimension $(3/2)(2n-1) - 1/2 = 3n-2$, and hence we find
perturbing irrelevant operators ${\cal O}_7$ and ${\cal O}_{10}$.

We can obtain mass terms for brane modes at the bottom of the
throat by substituting some of the $A$s and $B$s in each ${\cal
O}$ with their VEV, leaving a mass term (i.e. the operators are
``dangerously irrelevant'').  At the bottom of the throat these
VEVs are naturally of the scale $\anot \MMp$.  Hence a mass term
from of ${\cal O}_7$, will naturally scale like $m^2 \sim
(\anot)^5$, while a mass term from ${\cal O}_{10}$ behaves as $m^2
\sim (\anot)^8$.  Recalling that $z \sim a_0^3$, These are precisely the results for the leading (\ref{DistantFlux}) and subleading (\ref{O10DistantFlux})
perturbations from distant flux we found above, confirming from
the dual field theory point of view that these are the appropriate
corrections to the KS throat.

This analysis naturally suggests that it is possible to forbid the
larger mass term (\ref{DistantFlux}), leaving the smaller
(\ref{O10DistantFlux}) as the leading correction, by imposing a
discrete symmetry.  For example, $A_i \rightarrow - A_i$, $B_j
\rightarrow B_j$ is a symmetry of the KS field theory dual. It can
be mapped into a symmetry of the geometry as in \cite{KW}. Requiring that such a $Z_2$ symmetry can be extended to hold
throughout the geometry is enough to forbid (\ref{DistantFlux}).
It is easy to find examples of Calabi-Yau manifolds which admit
such a global $Z_2$ symmetry.

\sect{Computation of Effective Potential $V_{\rm eff}(X)$}

In this appendix we outline the derivation of eqn (\ref{supp}).
Consider $p$ particles on a sphere with radius $R$. We assume that
$p$ is large, and will work to leading order in $p$. Particle $i$
has a position ${\vec x}_i$ satisfying $|\vec x_i|^2 = R^2$. The
particles interact via a potential \be V(x_{ij}) = (x_{ij})^n \,
\qquad \quad x_{ij} = |{\vec x}_i - {\vec x}_j|. \ee Define $X$ as
the square root of the average (distance)${}^2$ between the
particles \be X^2 = {1\over p^2}\sum_{i\neq j} (x_{ij})^2 \ee Let
us assume that the motion of the particles is governed by the
Lagrangian \be L = {1\over 2} \sum_i \dot{x}^2_i - \sum_{i<j}
V(x_{ij}) - \sum_i \lambda_i(x_i^2 -R^2), \ee with corresponding
equation of motion \be \label{eom} \ddot {\vec x}_i = -
\sum_{k\neq i} \hat x_{ik} V'(x_{ik})   - 2 \lambda_i {\vec x}_i
\ee Starting with all particles at rest, we want to compute the
second time derivative $\ddot X$. We start from \ba X \ddot{X} \is
{1\over p^2} \sum_{i\neq  j} {\vec x}_{ij}\! \cdot (\ddot{\vec
x}_{i}-\ddot{\vec x}_j) \ea The plan is to evaluate the right-hand
side by inserting the equation of motion (\ref{eom}). This still
results in a complicated expression. However, we can simplify the
calculation by treating the Lagrange multipliers $\lambda_i$ in a
``mean field'' approximation, setting \be \lambda_i = \lambda_j =
\lambda \ee The mean field value is determined by the condition
that \ba \sum_i {\vec x}_i \! \cdot \ddot{\vec x}_i \is -
\sum_{i\neq j} {\vec x}_i\! \cdot \hat x_{ij} V'(x_{ij})
-  2 \lambda \sum_i x_i^2 \nonumber\\[2mm]
\is - {1\over 2} \sum_{i \neq j} {x}_{ij}\, V'(x_{ij})\;  - \; 2
\lambda\,p\, R^2 \, = \, 0 \ea We thus find \be \lambda = {n p V
\over 4 \, R^2}\, \qquad \quad V \equiv {1\over p^2} \sum_{i\neq
j} V(x_{ij}) \ee A straightforward calculation now gives \ba X
\ddot X \is {1\over p} \sum_{i \neq j} {x}_{ij}\, V'(x_{ij})\, -\,
{2 \lambda\over p^2} \sum_{i\neq j} x_{ij}^2
\nonumber\\[2mm]
\is \Bigl(1 - {X^2 \over 2 R^2} \Bigr) \, p \, n V \ea Identifying
$n V = X V'(X)$ gives equation (\ref{supp}).

\renewcommand{\Large}{\large}

\end{document}